\documentclass[10pt,article,oneside]{memoir}
\pdfoutput=1

\date{}
\counterwithout{section}{chapter}

\usepackage{amsthm, amssymb, amsmath, latexsym, amsfonts, mathcomp}
\usepackage{xfrac}
\usepackage{verbatim}
\usepackage{color}
\usepackage{mathrsfs}
\usepackage{listings}
\usepackage{array}
\usepackage{url}
\usepackage{footnote}
\usepackage[style=authortitle-ibid,url=true,doi=false,isbn=false]{biblatex}
\usepackage{microtype}
\usepackage{wrapfig}

\bibliography{Everything}
\usepackage{graphicx}

\allowdisplaybreaks

\newtheorem{proposition}{Proposition}[chapter]
\newtheorem{theorem}{Theorem}[chapter]
\newtheorem{lemma}{Lemma}[chapter]
\newtheorem{corollary}{Corollary}[chapter]
\newtheorem{definition}{Definition}[chapter]
\newtheorem{postulate}{Postulate}[chapter]
\newtheorem{axiom}{Axiom}[chapter]

\theoremstyle{definition}

\def\dd{\mathrm{d}}

\def\circle{S^1}

\def\R{\mathbb{R}}

\def\P{\mb{P}}                
\def\g{\mn{g}}                

\def\eps{\epsilon}

\def\Q{Q}
\def\P{P}

\def\g{\mathfrak{g}}

\newcommand{\dnabla}[1]{\ensuremath{\nabla_{\hspace{-0.02in} #1}}}

\DeclareCiteCommand{\parencite}[\mkbibparens]
  {\usebibmacro{prenote}}
  {\usebibmacro{citeindex}%
    \clearfield{url}%
    \usebibmacro{cite}}
  {\multicitedelim}
  {\usebibmacro{cite:postnote}}
\DeclareCiteCommand{\footcite}[\mkbibfootnote]
  {\usebibmacro{prenote}}
  {\usebibmacro{citeindex}%
    \clearfield{url}%
    \usebibmacro{cite}}
  {\multicitedelim}
  {\usebibmacro{cite:postnote}}

\DeclareCiteCommand{\parencitenn}[\mkbibparens]
  {\usebibmacro{prenote}}
  {\usebibmacro{citeindex}%
   \usebibmacro{citetitle}}
  {\multicitedelim}
  {\usebibmacro{cite:postnote}}

\DeclareCiteCommand{\citenn}
  {\usebibmacro{prenote}}
  {\usebibmacro{citeindex}%
   \usebibmacro{citetitle}}
  {\multicitedelim}
  {\usebibmacro{cite:postnote}}

\title{A Primer on Geometric Mechanics}

\author{\small{Christian Lessig}
        \\
        \small{Computational $\!+\!$ Mathematical Sciences, Caltech}
        }

\begin{document}

\maketitle

\vspace{-0.35in}
\begin{abstract} 
Geometric mechanics is usually studied in applied mathematics and most introductory texts are hence aimed at a mathematically minded audience. 
The present note tries to provide the intuition of geometric mechanics and to show the relevance of the subject for an understanding of ``mechanics''.
\end{abstract}


\section{How does geometry get into physics?}
Geometric mechanics employs modern geometry to describe mechanical systems. 
But how does geometry arise in mechanics?
For some common mechanical systems the space of all physically possible configurations is shown in Table~\ref{tab:math:gm:primer:config_space:examples}.
For a classical particle this is just Euclidean space since its state is completely described by its position, and for a pendulum every configuration is given by the angle $\theta$ with respect to a reference axis so that the circle $\circle$ provides the space of all possible configurations.
Already for the double pendulum, however, the situation becomes more interesting. 
Every configuration of the two arms is described by two angles, say $\theta$ and $\phi$, and since the arms are independent of each other the space of all possible states is $\circle \! \times \circle$.
But the tensor product $\circle \! \times \circle$ forms the torus $\mathbb{T}^2$.
Instead of the arms which represent the system in physical space, we hence have an alternative representation of the system where configurations are given by points $(\theta , \phi)$ on the ``doughnut'' with every point \emph{corresponding} to a displacement of the arms.
What is also apparent from the double pendulum is how constraints can be enforced intrinsically by choosing an appropriate geometric representation. 
The system could equivalently be described with the endpoints of the arms as particles in $\R^3$. 
But how many variables would then be needed? 
We would require two $3$-dimensional Cartesian coordinates to describe the positions and we would need two constraints, one for the plane the pendulum lies in and one for the unit spheres on which the particles move.
With the torus $\mathbb{T}^2 = \circle \! \times \circle$, the natural geometric structure of the double pendulum, the two angles $\theta$ and $\phi$ suffice to describe all configurations. 
We begin to see how geometry is an intrinsic part of mechanics and why the geometry should be respected: the space of all admissible configurations of a mechanical system has a natural geometric structure and constraints are intrinsically satisfied by the choice of the geometry. 
In more formal parlance, the configuration space $\Q$ of a mechanical system is a manifold, the generalization of a $2$-dimensional surface in space, and its topological and geometrical structure represent all physical states.

\begin{table}[ht!]
  \vspace{0.1in}
  \begin{center}
    \begin{tabularx}{1.0\textwidth}{X|X|X}
       & \hspace{0.15in} Physical System & \hspace{0.05in} Configuration space
      \\ \hline 
      \parbox[c]{1.2in}{\hspace{0.1in}Classical particle}
      &
      \parbox[c]{1em}{
      \centerline{\hspace{1.2in}
      \includegraphics[trim = 75mm 60mm 72mm 42mm, clip, scale=0.285]{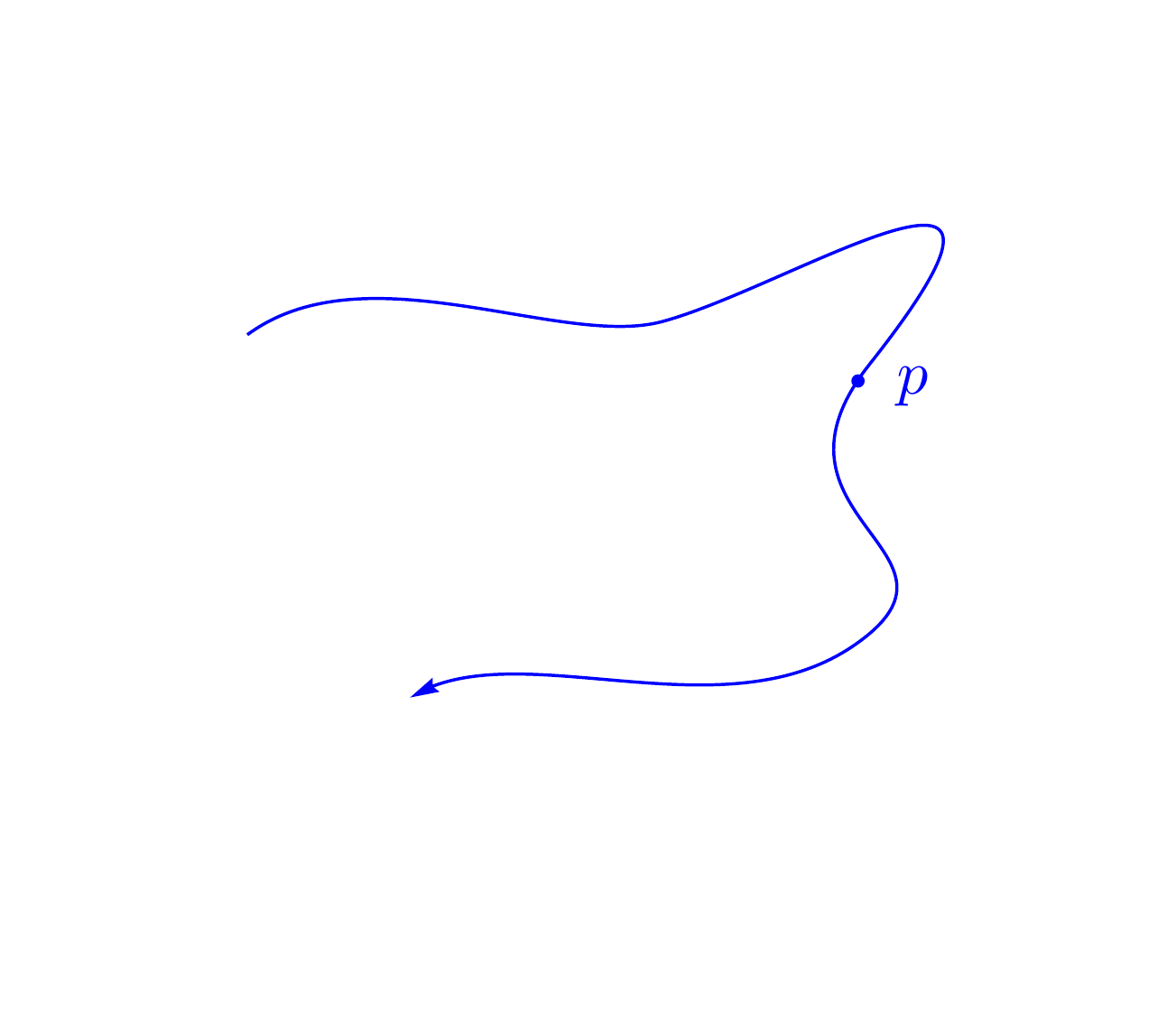}}
      }
      & 
      \parbox[c]{1em}{
      \centerline{\hspace{1.2in}
      \includegraphics[trim = 75mm 60mm 72mm 42mm, clip, scale=0.285]{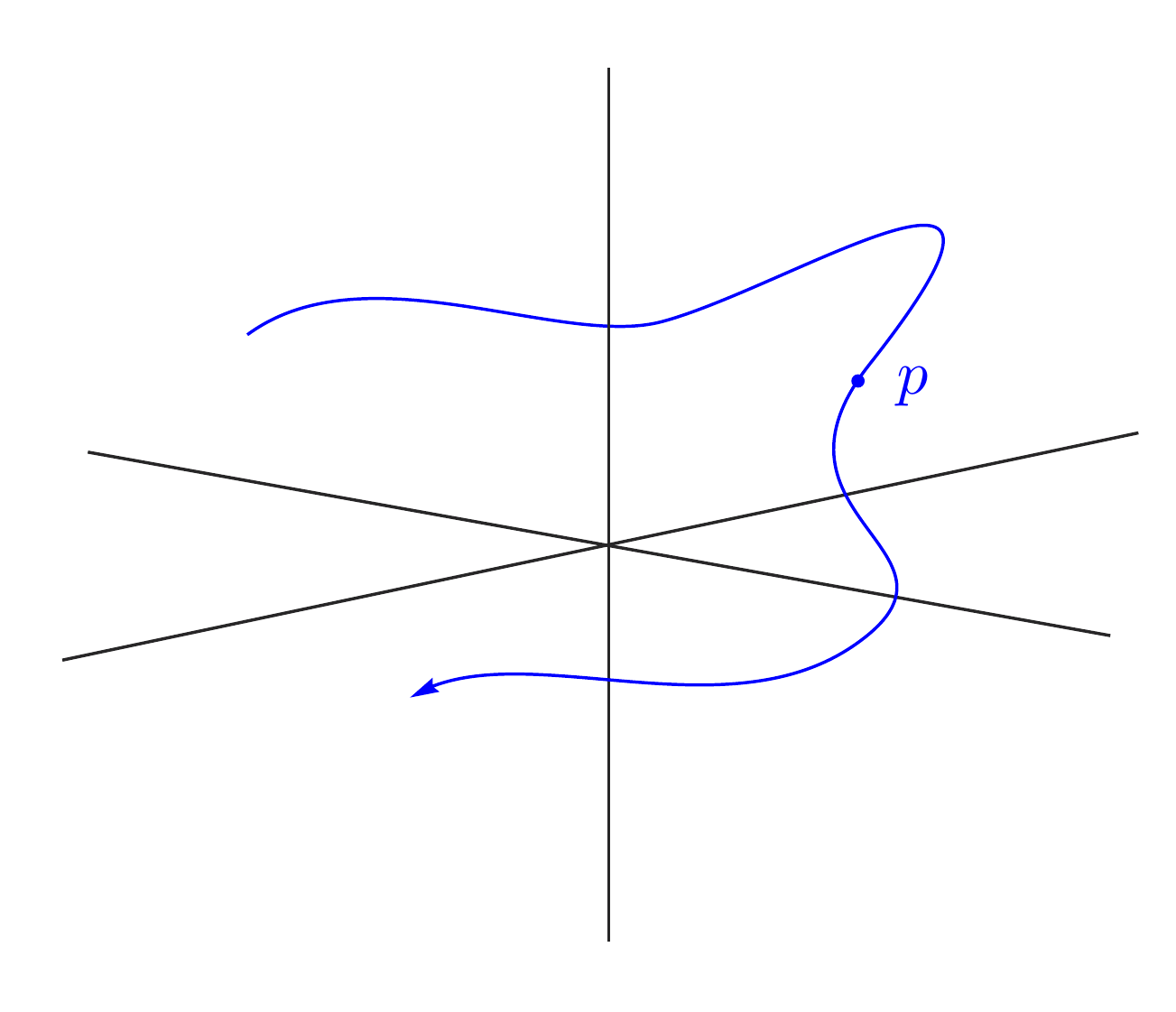}}
      }
      \\ \hline
      \parbox[c]{1.2in}{\hspace{0.1in}Single pendulum}
      &
      \parbox[c]{1em}{
      \centerline{\hspace{1.25in}
      \includegraphics[trim = 90mm 50mm 65mm 45mm, clip, scale=0.285]{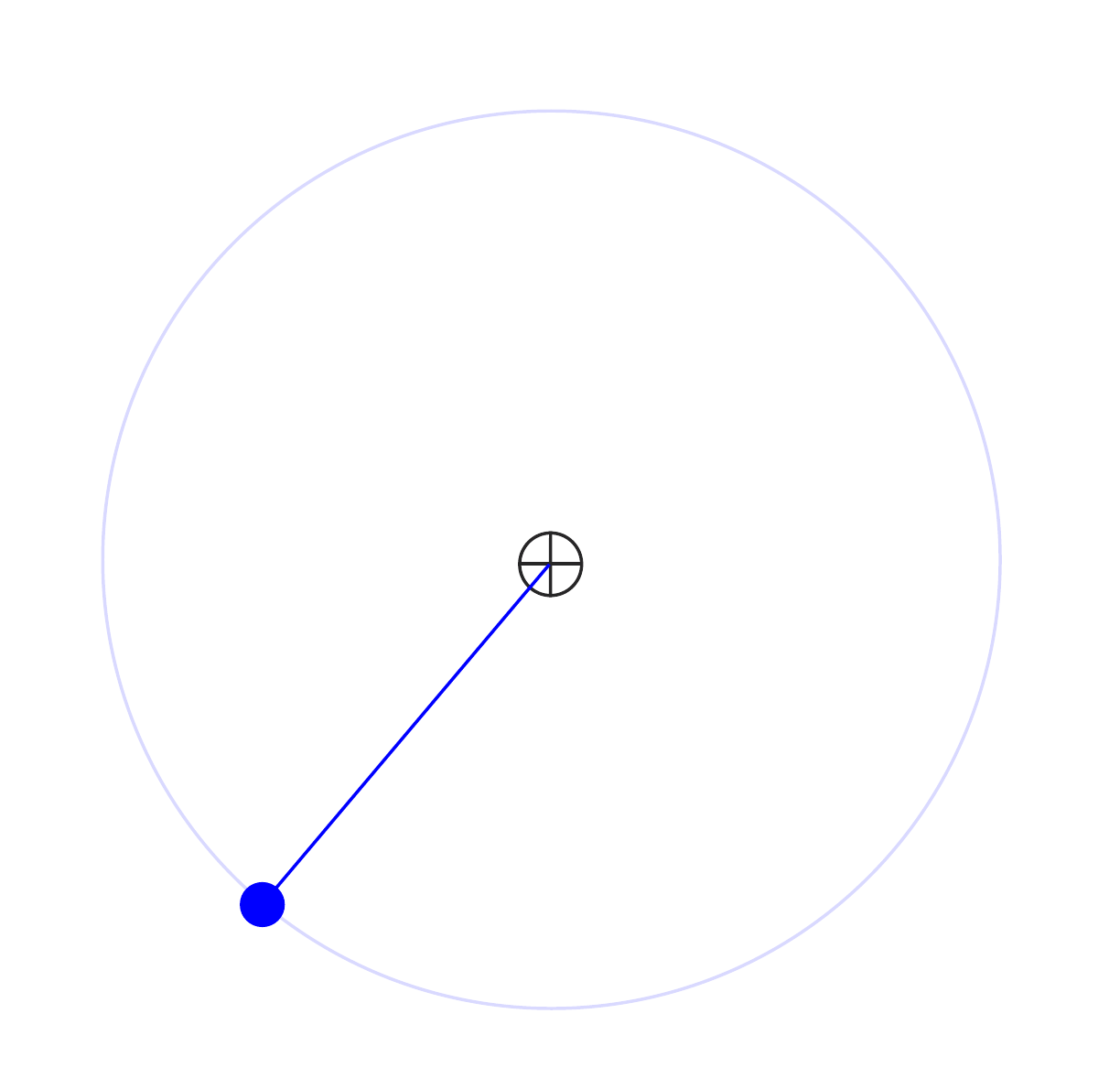}}
      }
      & 
      \parbox[c]{1em}{
      \centerline{\hspace{1.25in}
      \includegraphics[trim = 90mm 50mm 65mm 45mm, clip, scale=0.285]{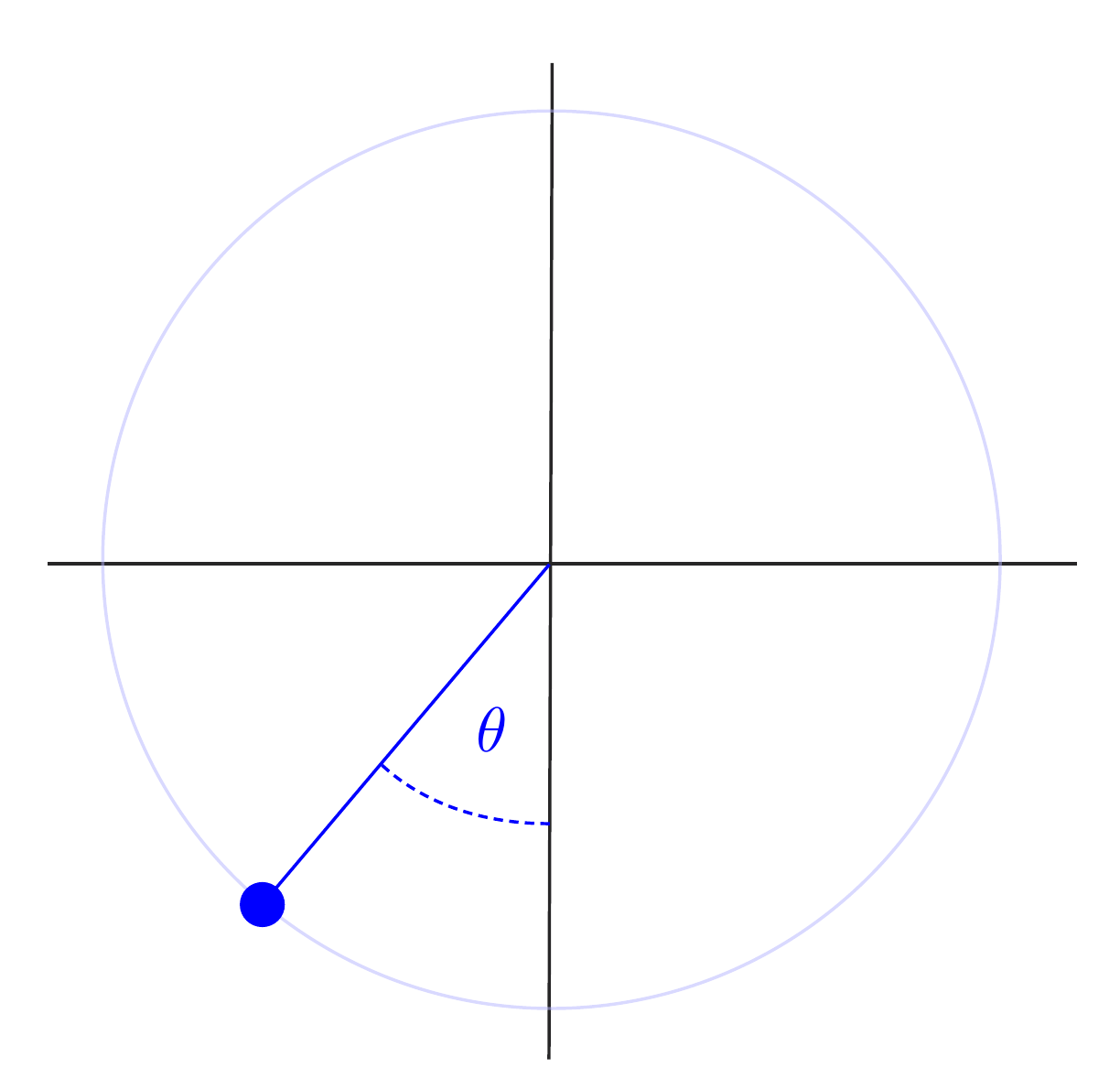}}
      }
      \\ \hline
      \parbox[c]{1.2in}{\hspace{0.1in}Double pendulum}
      & 
      \parbox[c]{1em}{
      \centerline{\hspace{1.25in}
      \includegraphics[trim = 90mm 50mm 65mm 45mm, clip, scale=0.285]{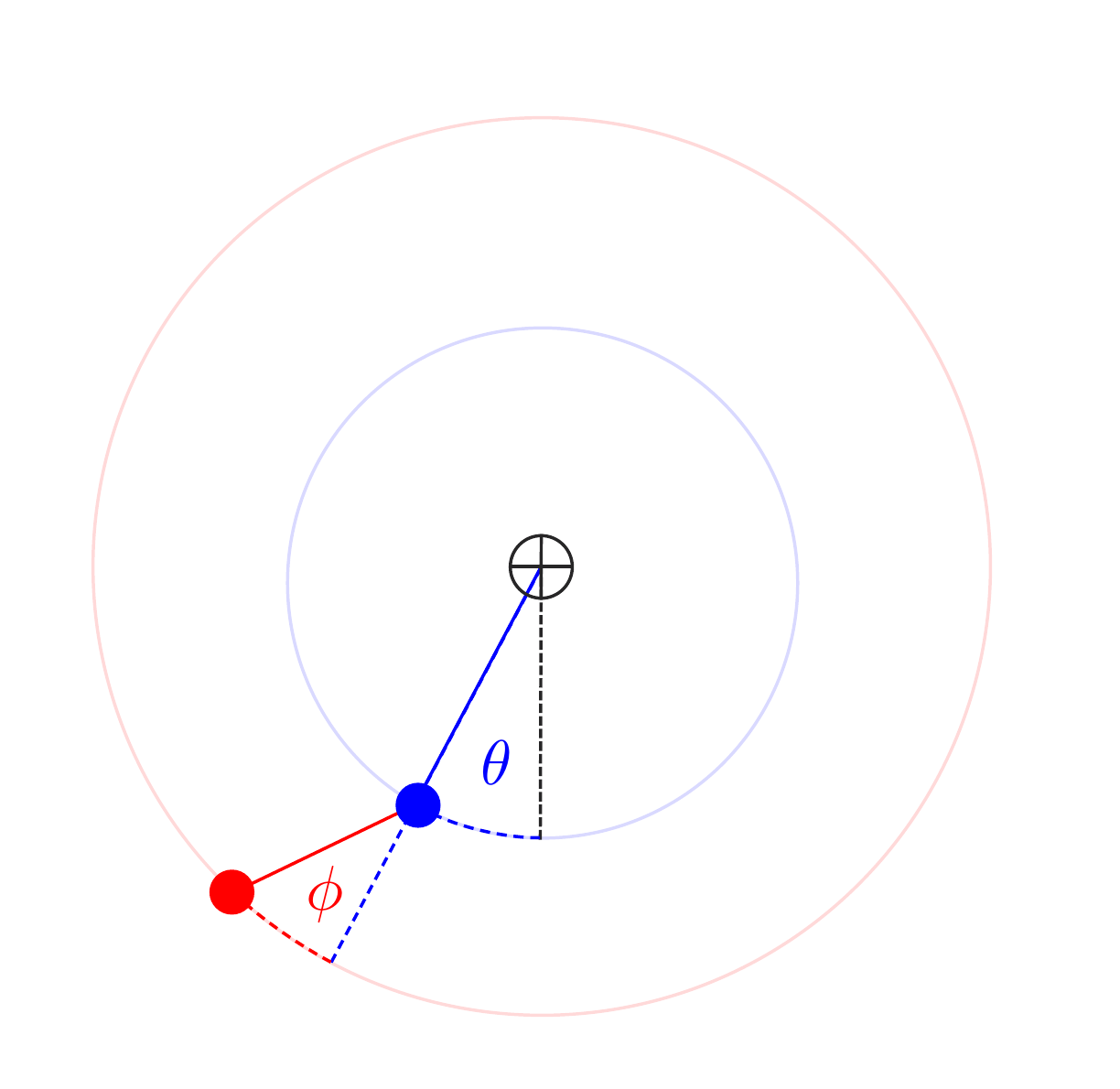}}
      }
      & 
      \parbox[c]{1em}{
      \centerline{\hspace{1.25in}
      \includegraphics[trim = 62mm 70mm 65mm 65mm, clip, scale=0.235]{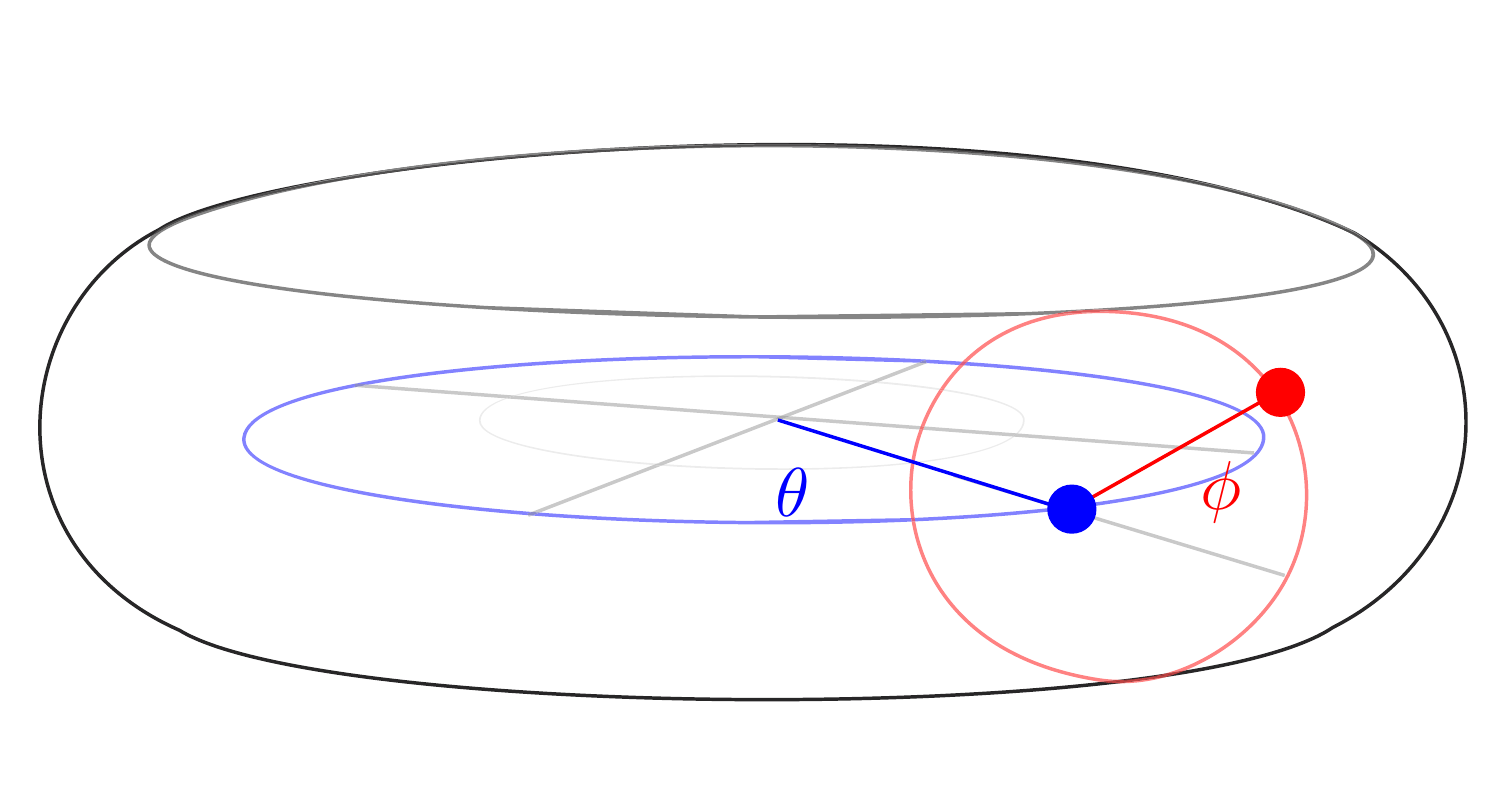}}
      }
      \\ \hline
      \parbox[c]{1.2in}{\hspace{0.1in}Euler top}
      &
      \parbox[c]{1em}{
      \centerline{\hspace{1.15in}
      \includegraphics[trim = 75mm 68mm 72mm 42mm, clip, scale=0.285]{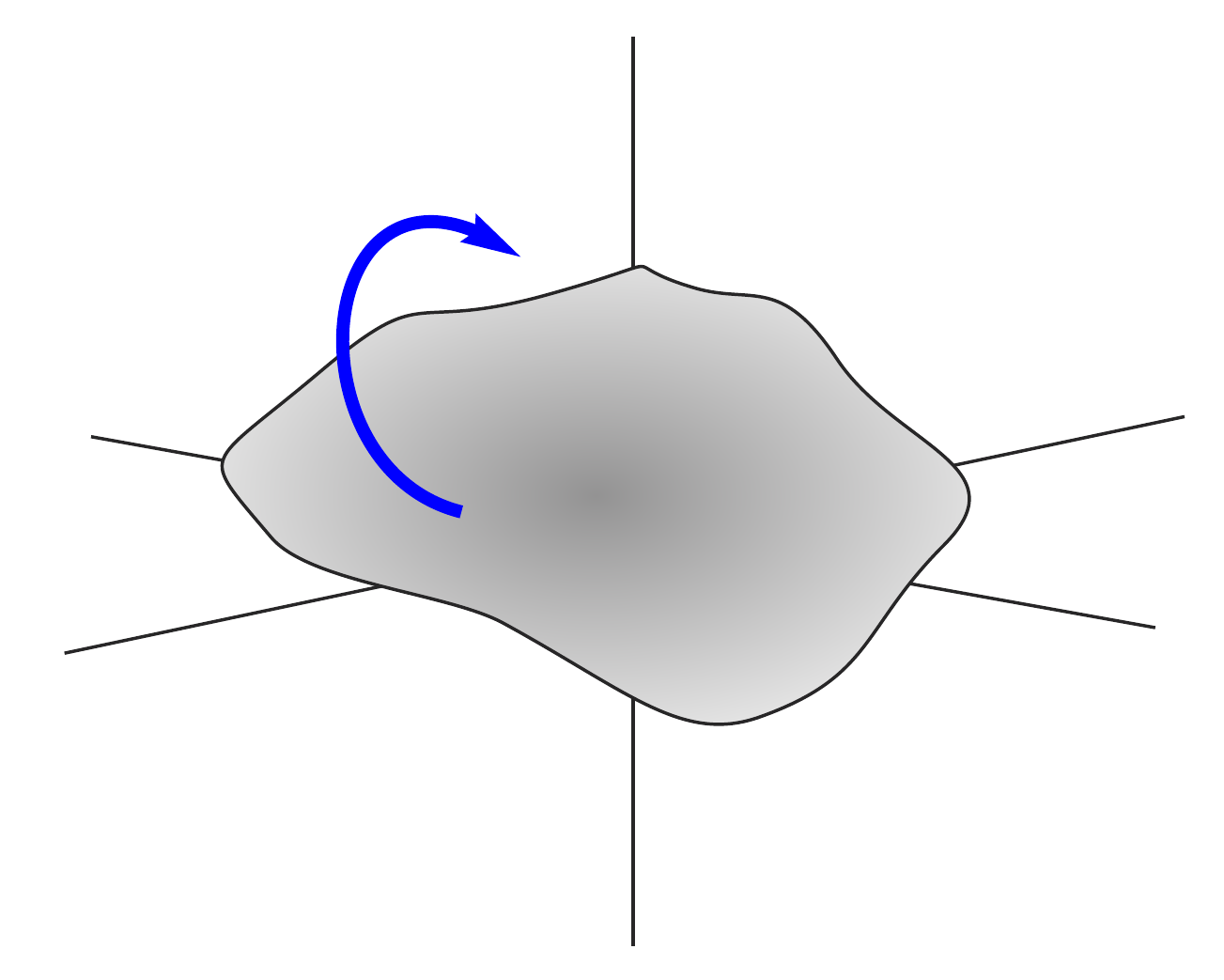}}
      }
      & 
      \parbox[c]{1em}{
      \centerline{\hspace{1.15in}
      \includegraphics[trim = 75mm 65mm 72mm 45mm, clip, scale=0.285]{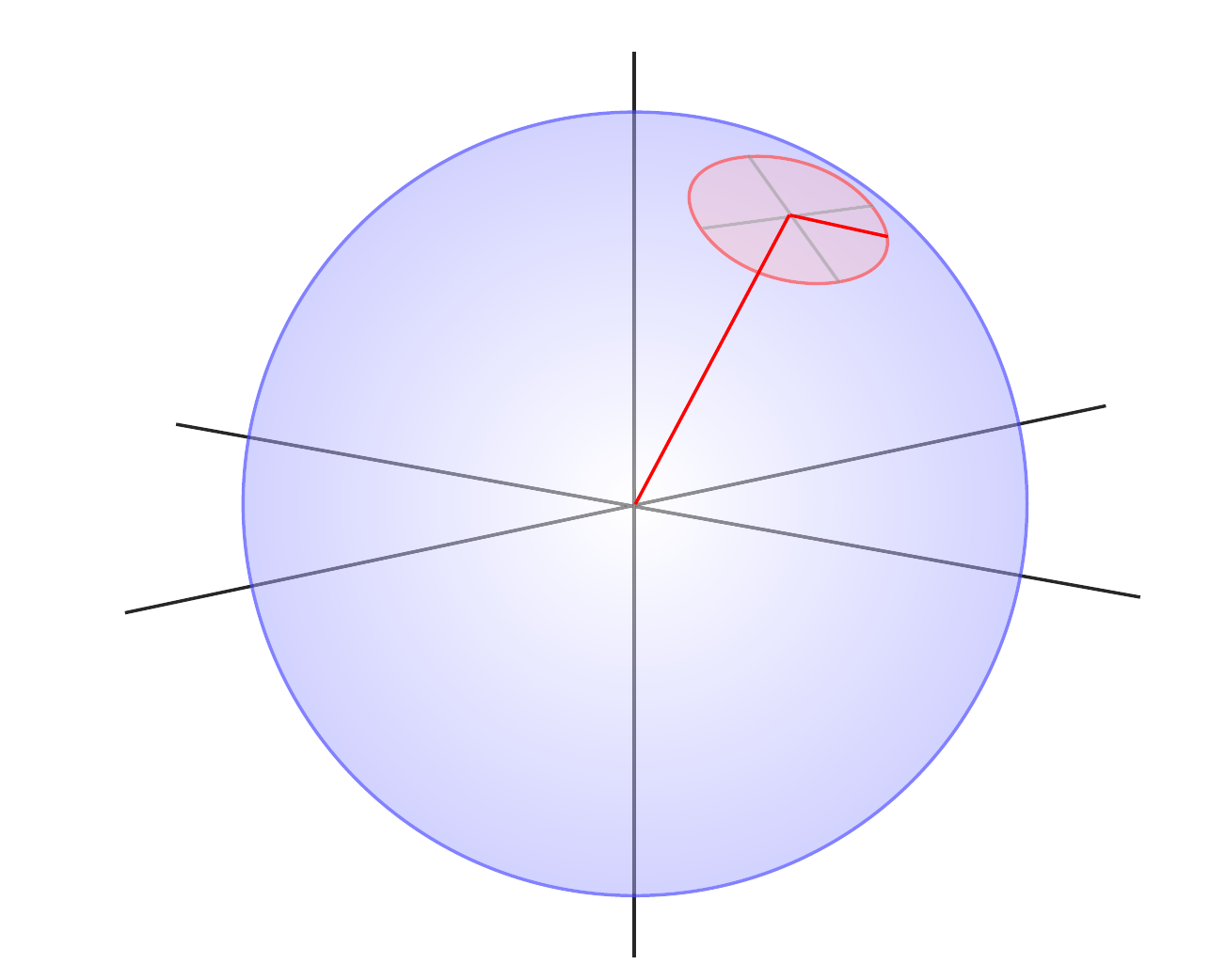}}
      }
      \\ \hline
      \parbox[c]{1.2in}{\hspace{0.1in}Ideal Euler fluid}
      & 
      \parbox[c]{1em}{
      \centerline{\hspace{1.25in}
      \includegraphics[trim = 75mm 60mm 65mm 40mm, clip, scale=0.26]{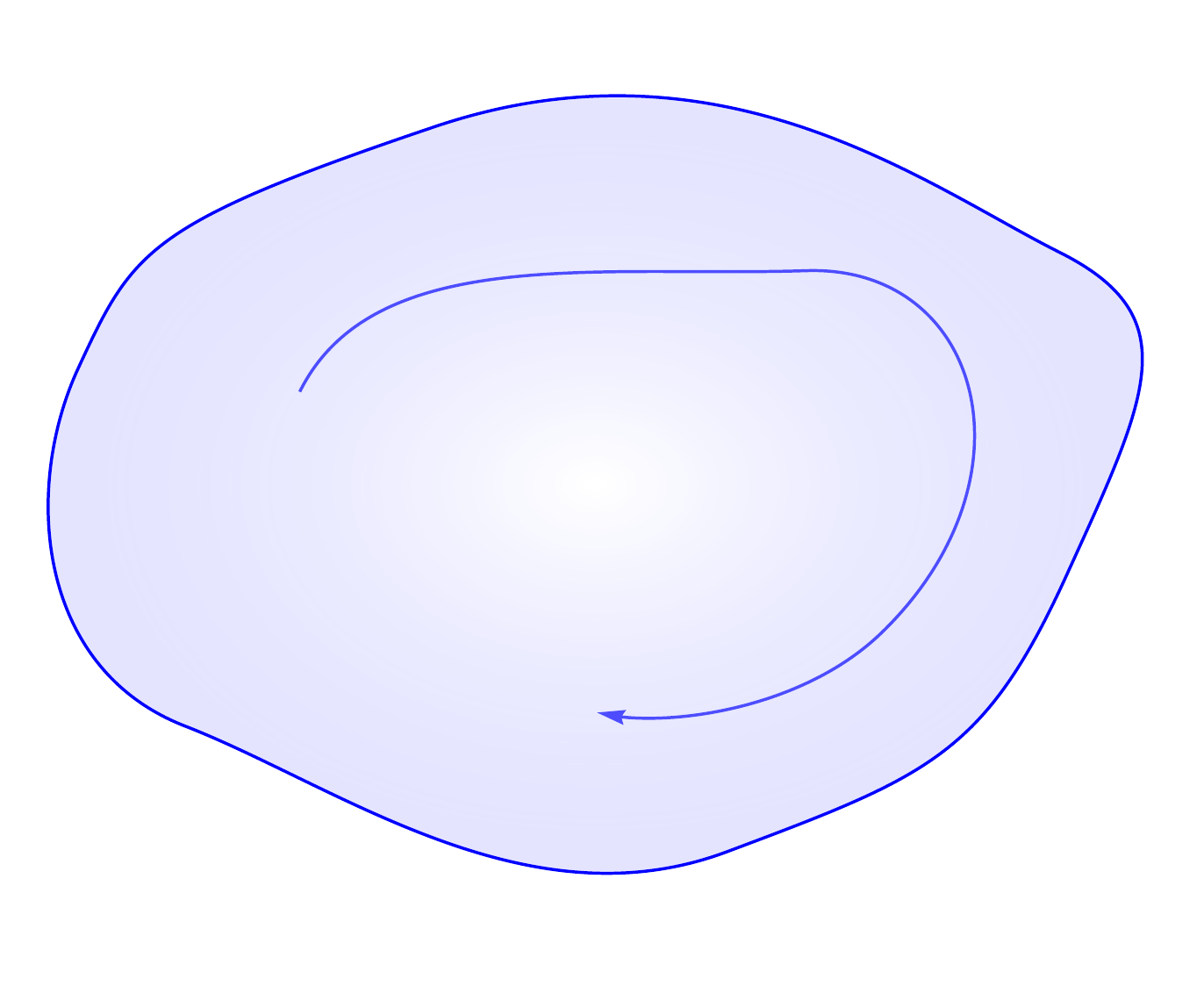}}
      }
      & 
      \parbox[c]{1em}{
      \centerline{\hspace{1.25in}
      \includegraphics[trim = 38mm 70mm 27mm 70mm, clip, scale=0.18]{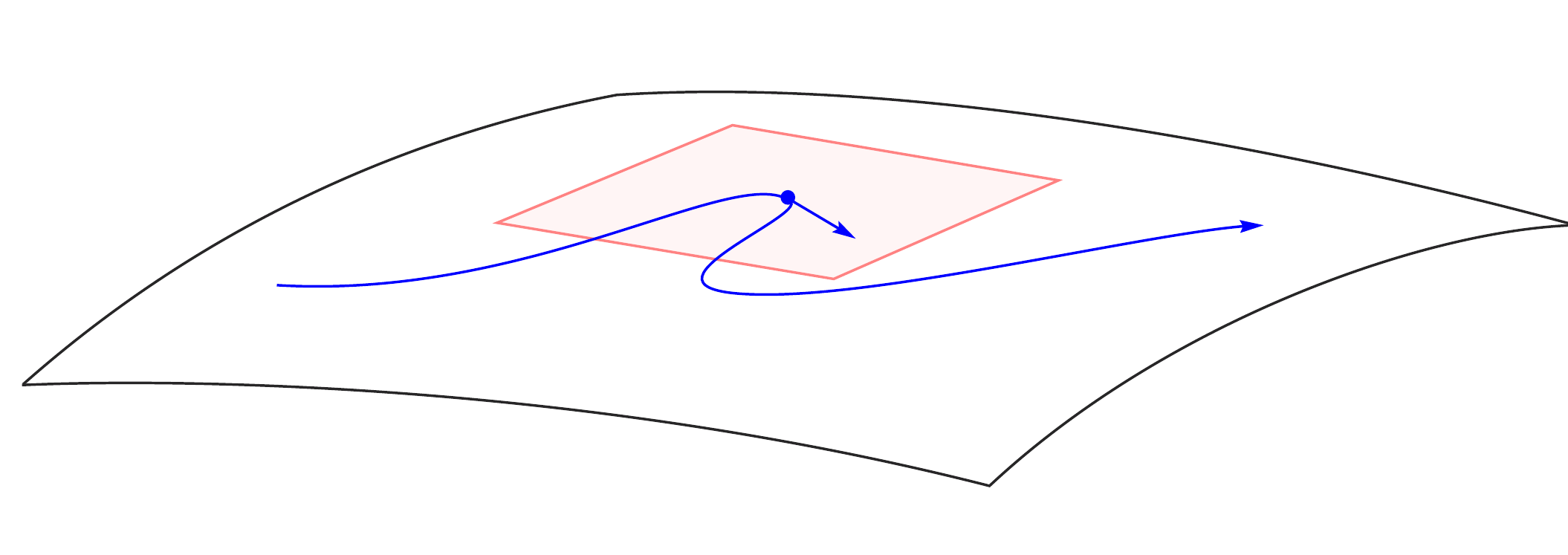}}
      }
    \end{tabularx}
  \end{center}
  \caption{Configuration space, the space of all physically valid states, for some classical mechanical systems.}
    \label{tab:math:gm:primer:config_space:examples}
\end{table}

The description of the configurations of a system as points on a manifold is the principal premise of geometric mechanics and it enables to illustrate the system's structure even when the configuration space is complicated and abstract, cf. again Table~\ref{tab:math:gm:primer:config_space:examples}, providing the inherent intuition of geometric mechanics.

\begin{figure}
  \setlength{\abovecaptionskip}{-20pt}
  \begin{center}
    \centerline{\hspace{-0.1in}
    \includegraphics[trim = 28mm 92mm 20mm 80mm, clip, scale=0.53]{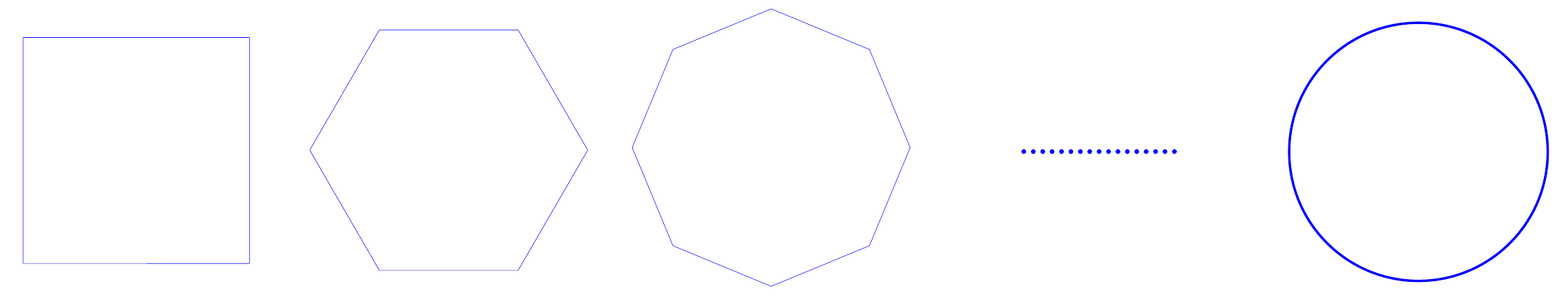}}
  \end{center}
  \caption{Continuous rotational symmetry as the limit of discrete rotational symmetries. Rotating a regular $n$-gon by $2 \pi / n$ yields the same $n$-gon---it is preserved under the discrete set of rotations. In the continuous limit the rotation by an arbitrary angle preserves the circle $\circle$.}
  \label{fig:math:gm:primer:continuous_symmetry}
\end{figure}

\section{What have a butterfly and a stone in common?}
We could end here, with manifolds as configuration spaces, and we would obtain a rich and vigorous theory.
However, many mechanical systems have another and complementary geometric structure: symmetry.\footnote{For a discussion of the history of the concept and its important in modern physics see~\parencite{Brading2007}.} 
In contrast to the discrete symmetries that might come to one's mind, such as the mirror symmetry of a butterfly or the discrete radial symmetry of flowers, mechanical system have continuous symmetries as depicted in Fig.~\ref{fig:math:gm:primer:continuous_symmetry}.
For example, for the classical particle considered before we can translate the coordinate system without affecting its motion---as two mirror images are equivalent so is a particle described in a translated reference frame.
The single pendulum possesses a rotational symmetry where we can rotate the reference axis without changing its physical behaviour, and a similar symmetry also exists for the Euler top: a rigid body, such as a stone, that is fixed in space but free to rotate around any axis, see again Table~\ref{tab:math:gm:primer:config_space:examples}.\footnote{It is named Euler top since it was Euler who first wrote down the correct equations of motion.}
At any time, the configuration of the Euler top is described by a rotation with respect to an initial configuration---whose geometric structure is illustrated nicely by the Poincar{\'e} map---and the continuous symmetry hence arises again from the arbitrariness of the reference configuration.
Going from the $1$- and $3$-dimensional symmetries of the pendulum and the Euler top to an ``infinite'' dimensional rotational symmetry leads to the geometric structure of an ideal Euler fluid.\footnote{This analogy was first pointed out by Arnold~\parencitenn{Arnold1966}.} 
For this system, the geometry is too complicated to be visualized directly and we have to resort to an iconic representation as in Table~\ref{tab:math:gm:primer:config_space:examples}.
However, all configurations can again be described with respect to an initial reference configuration by considering the trajectories traced out by the fluid ``particles'', and globally this is represented by a diffeomorphism which is volume preserving since the fluid is incompressible.\footnote{For a formal definition of a diffeomorphism see Definition~\ref{def:math:diffeomorphism} and also Chapter~\ref{sec:math:gm:lie_groups:infinite}. Intuitively, one can think of a diffeomorphism as a smooth map between continuous domains that has a smooth inverse, that is every point in the first domain is smoothly mapped to a point in the second domain, and there is a map that ``reverses'' going from the first to the second domain.} 

Formally, continuous symmetries are described by Lie groups and their action on configuration space, and their importance lies in the conserved quantities and reduced descriptions which arise from them.
We will consider these aspects in more detail in the following when we discuss dynamics.

\begin{figure}[t]
  \begin{center}
    \includegraphics[trim = 25mm 82mm 25mm 70mm, clip, scale=0.5]{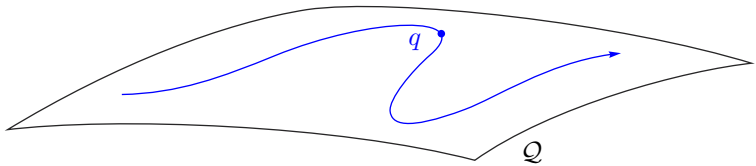}
  \end{center}
  \caption{Configuration space $\Q$ of physical system with each point $q \in \Q$ being a valid physical configuration of the system. Time evolution corresponds to a curve $q(t)$ on the manifold $\Q$.}
  \label{fig:math:gm:primer:config_space:iconic}
\end{figure}

\section{How does a rubber band describe the dynamics of a rigid body?}
\begin{wrapfigure}{r}{5.8cm} 
  \setlength{\abovecaptionskip}{-0pt}
  \vspace{-0.12in}
  \centering
  \centerline{
  \hspace{-0.1in}
  \includegraphics[trim = 63mm 73mm 60mm 75mm, clip, scale=0.4]{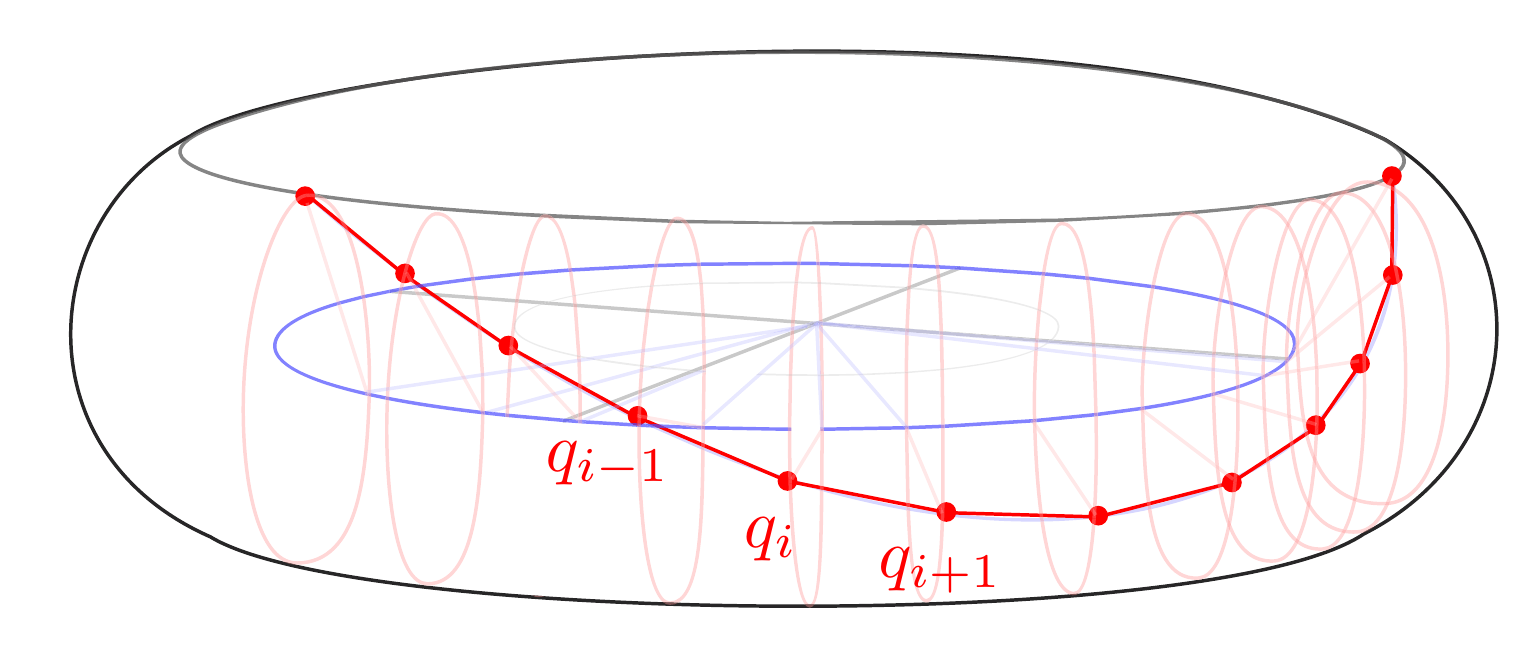}}
  \caption{}
  \vspace{-0.2in}
  \label{fig:math:gm:primer:torus:time_evo}
\end{wrapfigure}

We already learned that the configurations of a mechanical system are naturally described by points on its configuration manifold $Q$. 
What we are really interested in, however, are not isolated configurations but time evolution. 
\begin{figure}[t]
  \begin{center}
    \includegraphics[trim = 25mm 82mm 25mm 70mm, clip, scale=0.5]{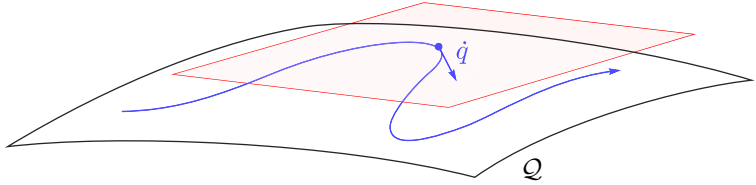}
  \end{center}
  \caption{The velocity $\dot{q}(t)$ of the curve $q(t)$ lies in the tangent bundle $T_q Q$ of the configuration manifold $Q$.}
  \label{fig:math:gm:primer:config_space:velocity}
\end{figure}
A set of consecutive states at times $t_1 < t_2 < \ldots$ is thus given by a sequence $\left( q(t_1), q(t_2), \ldots \right)$ of locations on $\mathbb{T}^2$, see Fig.~\ref{fig:math:gm:primer:torus:time_evo}. 
To understand how this is described in geometric mechanics let us consider again the double pendulum.
Each configuration of the system is given by two angles which represent a point $q = (\theta,\phi)$ on the torus $\mathbb{T}^2$.
When the time interval $\Delta t_i$ between configurations $q(t_i)$ and $q(t_{i+1})$ becomes vanishingly small, we surely expect that also the distance 
between the points on the torus goes to zero. 
But then the configurations have to form a smooth curve $q(t)$, and since the $q(t_i)$ lie on the configuration space of the double pendulum it is a curve on the torus $\mathbb{T}^2$.
A little thought shows that our reasoning was independent of the chosen example and that for any system the time evolution can be described by a curve $q(t) : [a,b] \to Q$ on configuration space.\footnote{For general physical systems it is not necessarily satisfied that the curve on configuration space is smooth. For example, for systems with impact the curve is in general only continuous. However, in these introductory notes we will restrict ourselves to smooth curves}
Hence, in geometric mechanics much intuition also exists for the time evolution of a system and we can illustrate it as a ``marble'' tracing out its path on configuration space.
When configuration space is too complex to be visualized directly, such as for the Euler fluid, we will retain the geometric intuition again by considering a curve on an iconic representation as depicted in Fig.~\ref{fig:math:gm:primer:config_space:iconic}. 

\begin{wrapfigure}{l}{5.8cm} 
  \setlength{\abovecaptionskip}{-0pt}
  \vspace{-0.12in}
  \centering
  \centerline{
  \hspace{-0.1in}
  \includegraphics[trim = 63mm 73mm 60mm 75mm, clip, scale=0.4]{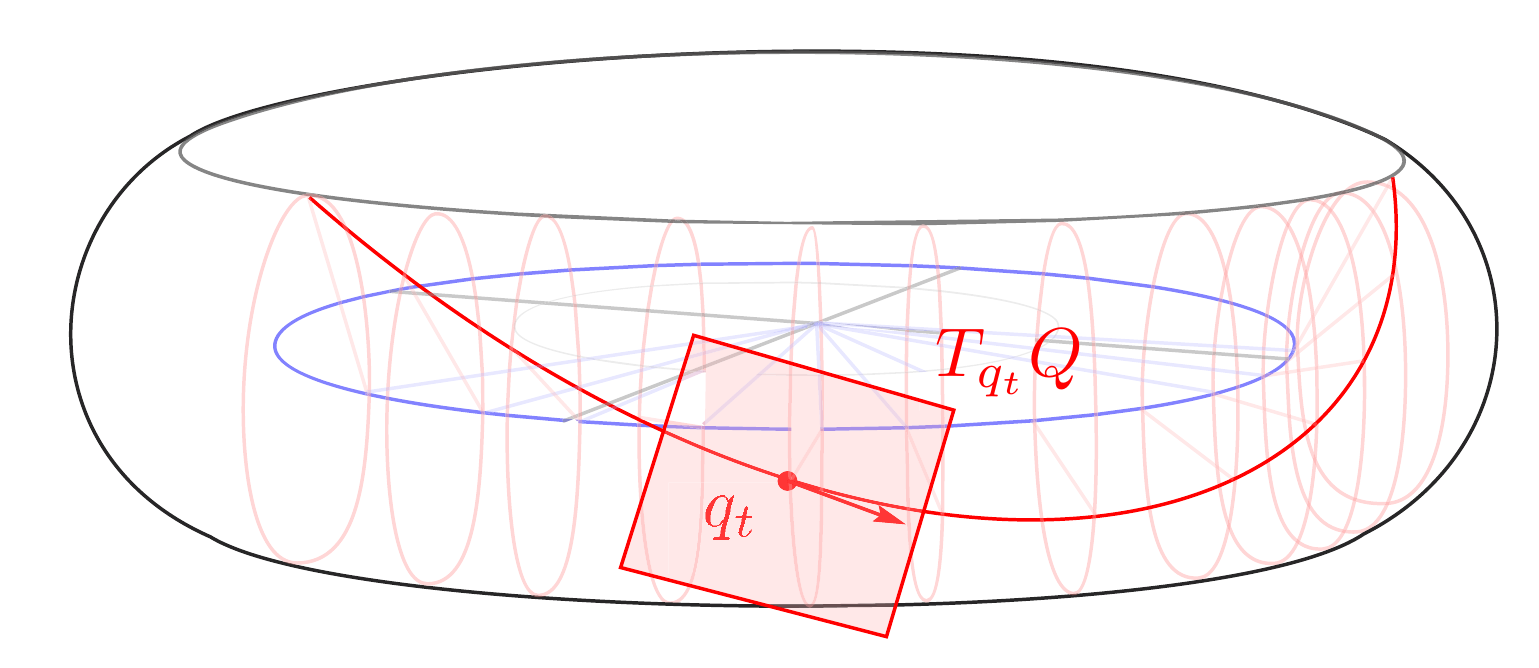}
  }
  \caption{}
  \label{fig:math:gm:primer:torus:velocity}
  \vspace{-0.2in}
\end{wrapfigure}

Given a curve $q(t) : [a,b] \to Q$ describing the time evolution of a system, we know its configuration for all times $t \in [a,b]$. 
Unfortunately, we rarely have this information at our disposal and all we know in most instances is the system's current state---although we still would like to determine its future configurations. 
Even worse, the present configuration is usually not even sufficient to determine the time evolution.
Fortunately, for many systems knowing its position $q(t)$ and its velocity $\dot{q}(t) = v(t) = d q(t) / dt$ provides the needed information.
The geometry of the velocity vector $\dot{q}(t)$ is easily understood when we consider again the double pendulum and assume that the second pendulum is at rest, in which case the time evolution is described by a curve $q(t)$ on the equator of the torus $\mathbb{T}^2$.
The velocity vector $\dot{q}(\bar{t})$ for some time $\bar{t}$ is by construction tangent to the curve.
But since $q(t)$ lies on the torus $\mathbb{T}^2$, the tangent $\dot{q}(\bar{t})$ lies also in the tangent space of the manifold at the point $q(\bar{t})$ along the curve, see Fig.~\ref{fig:math:gm:primer:torus:velocity}.\footnote{As usual, the tangent space can be interpreted as the best linear approximation to the manifold at a point.}
It is again easy to see that there was nothing special about our example, and that for any system the velocity $\dot{q}(t)$ is tangent to the curve $q(t)$ describing the time evolution of the system and it lies in the tangent space $T_{q(t)} Q$ of configuration space at $q(t)$, see Fig.~\ref{fig:math:gm:primer:config_space:velocity} for the general picture one should have in mind.

So far we discussed how curves on configuration space enable to describe the states of a system over time. 
However, usually we are only given the current configuration and what we are interested in is the state in the future.
\begin{figure}[t]
  \begin{center}
  \includegraphics[trim = 25mm 82mm 25mm 70mm, clip, scale=0.5]{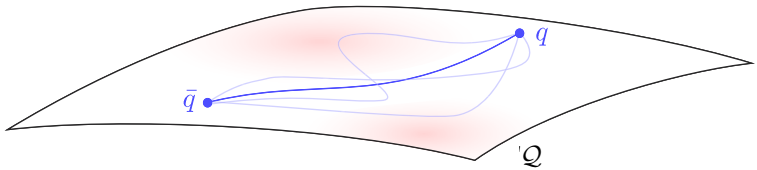}
  \end{center}
  \caption{A physical path between two states $\bar{q}$ and $q$ is a stationary point of the action (blue) in the space of all possible paths on configuration space $Q$.}
  \label{fig:math:gm:primer:lagrangian:action}
\end{figure}
A recipe for obtaining a description of the time evolution is provided by Hamilton's principle of least action.
Intuitively, it states that physical paths on configuration space are paths of least resistance with respect to an action functional $S(q(t))$---or the paths a rubber band would settle in on configuration space around the ``hills'' defined by the Lagrangian $L(q(t),\dot{q}(t))$, see Fig.~\ref{fig:math:gm:primer:lagrangian:action} for the geometric intuition.\footnote{The idea of least resistance is even more apparent in the Gauss-Hertz principle of least curvature~\parencitenn{Hertz1894} but it is technically more involved and less general than Hamilton's principle.} 
Formally, the action principle is given by
\begin{align*}
  0 
  = \frac{\delta S}{\delta q(t)} 
  &= \frac{d}{d \eps} S(q(t) + \eps \, r(t)) ,
\end{align*}
where the action $S$ corresponds to the energy along the rubber band as defined by the Lagrangian
\begin{align*}
  0 
  = \frac{\delta S}{\delta q(t)}
  &= \frac{d}{d \eps} \int L \left( q(t) + \eps \, r(t) , \frac{d}{dt} ( q(t) + \eps \, r(t) ) \right) dt ,
\end{align*} 
and it states that the paths on configuration space taken by a physical system correspond to stationary points $\delta S / \delta q(t) = 0$ of the action functional $S(q(t))$ where the functional derivative $\delta S / \delta q(t)$ vanishes.
Physical trajectories are hence the local extrema of the action $S(q(t))$, analogous to the local extrema of a function over the real line which are the stationary points of the ordinary derivative.
The Lagrangian $L(q(t),\dot{q}(t))$ can be understood as the characteristic function of a system---depending both on its configuration $q(t)$ and its velocity $\dot{q}(t)$---and it is usually defined as the kinetic minus the potential energy of a system.
For example, for a classical particle of mass $m$ in a potential $V(q)$ the Lagrangian is
\begin{align*}
  L(q,\dot{q}) = \frac{m}{2} \Vert \dot{q} \Vert^2 - V(q) .
\end{align*}
Using the calculus of functional derivatives, one can derive from Hamilton's principle differential equations describing the motion of the system.
In the general case, the equations are known as Euler-Lagrange equations, and for the above Lagrangian these are equivalent to Newton's equations of motion. 

Next to the action principle, an alternative way to describe the dynamics of a mechanical system is Hamiltonian mechanics where the time evolution is governed by the system's energy.
Instead of using the velocity $\dot{q} \in T Q$ which determines the change in a system's configuration, it is then useful to employ a description of the change in the system's (kinetic) energy $T$.
This is conveniently expressed using a co-vector $p$ in the dual space $T^*Q$ ``measuring'' the change $\delta T$ with the pairing $p(\dot{q}) = p \cdot \dot{q}$.
\begin{figure}[t]
  \begin{center}
    \includegraphics[trim = 25mm 82mm 25mm 70mm, clip, scale=0.5]{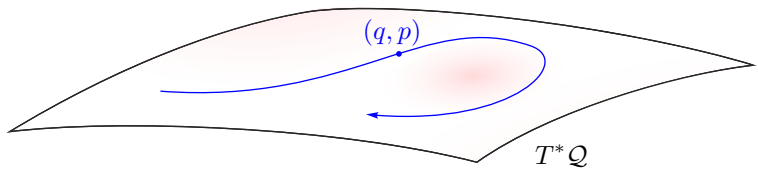}
  \end{center}
  \caption{In Hamiltonian mechanics the time evolution of a mechanical system is considered on phase space $\P = T^*Q$ and governed by the Hamiltonian (light red) which for each phase space point $(q,p)$ defines the total energy of the system.}
  \label{fig:math:gm:primer:hamiltonian:phase_space}
\end{figure}
For the classical particle we considered before, the co-vector is given by $p = m \, \dot{q}$ where the mass $m$ expresses the dependence on the kinetic energy $T = \sfrac{1}{2} \, m \, \Vert \dot{q} \Vert^2$, and $p$ is hence what Leibniz and Newton called momentum.\footnote{Incidentally, we also have $p(\dot{q}) = p \cdot \dot{q} = m \, \dot{q} \cdot \dot{q} = m \, \Vert \dot{q} \Vert^2 = 2 \, T$. But didn't we say the momentum determines the \emph{change} in the kinetic energy? How can $p$ at the same time provide the change \emph{and} the total energy?}
With the momentum, an alternative description of a system is possible by ``lifting'' it to the cotangent bundle $T^*Q$ and on this 'phase space' a configuration is then given by the system's position $q$ as well as its momentum $p$, see Fig.~\ref{fig:math:gm:primer:hamiltonian:phase_space}. 
Evidently, the lift makes a system's description more complex in that the number of variables describing a configuration is doubled.
However, as might already appear reasonable from intuition, it enables a ``simpler'' description of the dynamics.
Using a similar reasoning as before, it is not hard to see that on $T^*Q$ time evolution is again represented by a smooth curve $z(t) = (q(t),p(t)) : [a,b] \to T^*Q$.
However, in contrast to Lagrangian mechanics on configuration space where Hamilton's action principle determines trajectories, the evolution of a point $z(t) = (q(t),p(t))$ along the curve on phase space is governed by a simple law: the total energy of the system given by the Hamiltonian $H(q,p) : T^* \Q\to \R$ is conserved.\footnote{For some systems the Hamiltonian does not directly represent the total energy of a system but these are far beyond the scope of our considerations.} 
This leaves us again with the question how we can determine the curve describing the time evolution from an initial configuration?
To find the answer, let us return to the classical particle in a potential.
The Hamiltonian for the system is given by 
{\setlength{\belowdisplayskip}{6pt}\setlength{\abovedisplayskip}{6pt}
\begin{align*}
  H(q,p) = \frac{\Vert p \Vert^2}{2 \, m} + V(q)
\end{align*}}%
and its phase space is $\R^3 \times \R^3$ since the tangent space of Euclidean space $\R^3$ is the space itself and since we can think of the momentum as a vector in $\R^3$.\footnote{In most applications it is useful to \emph{not} identify $T\R^n$ with $\R^n$ but to carefully distinguish the two spaces, and it is similarly usually not advisable to identify $TQ$ and its dual $T^*Q$ even if this possible using a metric. However, for the simplicity of our argument we will employ the usual obfuscations.}
But in Euclidean space it is not hard to find the direction where a function $f : \R^3 \to \R$ does not change: the gradient $\nabla f$ determines the direction of maximal change and the change in an arbitrary direction $\vec{u}$ is given by the dot product $\nabla f \cdot \vec{u}$, and hence any vector orthogonal to the gradient direction $\nabla f$ defines a direction where the value of $f$ is conserved.
\begin{figure}[t]
  \begin{center}
    \includegraphics[trim = 25mm 82mm 25mm 70mm, clip, scale=0.5]{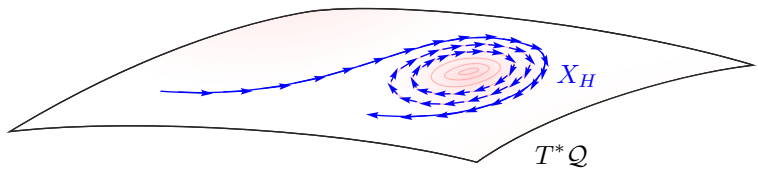}
  \end{center}
  \caption{Flow on phase space along the Hamiltonian vector field $X_H$ (blue arrows) preserving the level sets of the Hamiltonian (light red).}
  \label{fig:math:gm:primer:hamiltonian:flow}
\end{figure} 
With the gradients with respect to position and momentum being $\dnabla{q}$ and $\dnabla{p}$, respectively, the tangent vector $X_H$ on phase space defining the direction where the Hamiltonian does not change is given by 
\begin{align*}
  X_H = \mathbb{J} \, \dd H 
  \cong
  \left( \! \! \begin{array}{c}
    \dot{q} \\ \dot{p} 
  \end{array} \! \! \right)
  \!  = \!
  \left( \! \begin{array}{cc}
    0 & I \\ -I & 0
  \end{array} \! \right) \!
  \left( \! \! \begin{array}{c}
    \dnabla{q} \, H \\ \dnabla{p} \, H
  \end{array} \! \! \right)
  = 
  \left( \! \! \begin{array}{c}
    \ \ \dnabla{p} \, H \\ - \dnabla{q} \, H
  \end{array} \! \! \right)
\end{align*}
where $I$ denotes the $3 \times 3$ identity matrix and the above equations are known as Hamilton's equations.
The symplectic matrix $\mathbb{J}$ indeed ensures that the Hamiltonian $H(q,p)$ is conserved along the flow of the Hamiltonian vector field $X_H$ since 
\begin{align*}
  \dd H \left( X_H \right) 
  = \nabla H \cdot X_H
  = \dnabla{q} \, H \cdot \dnabla{p} \, H - \dnabla{p} \, H \cdot \dnabla{q} \, H = 0  
\end{align*}
and although the flow is defined on $6$-dimensional phase space much intuition about its behaviour can be gained by considering the underlying geometry as depicted in Fig.~\ref{fig:math:gm:primer:hamiltonian:flow}.
Needless to say, the above conception of the dynamics of Hamiltonian mechanics applies to any phase space when a generalized gradient and an intrinsic definition of the symplectic matrix are employed.

Lagrangian and Hamiltonian mechanics provide alternative descriptions of a mechanical system, and when they are equivalent, as is usually the case, one can change the point of view using the Legendre transform.
However, each perspective also provides its own merits and demerits, and often one of the descriptions appears more natural.
For example, as we saw before, Lagrangian mechanics is defined on configuration space $Q$ while Hamiltonian mechanics employs ``lifted'' dynamics on phase space $\P = T^* Q$.
This leads to second order differential equations for the time evolution in the Lagrangian picture and to first order equation in the Hamiltonian; more concretely, Newton's equations, which are equivalent to the Euler-Lagrange equation for a Lagrangian of the form considered before, depend on the acceleration, the second derivative of position with respect to time, while in Hamilton's equations as introduced above the Hamiltonian vector field $X_H$ is the first time derivative of position and momentum---it is this reduction from second to first order differential equations which provides the ``simplification'' in Hamiltonian dynamics which we advertised before.

\begin{figure}[t]
  \setlength{\abovecaptionskip}{-10pt}
  \begin{center}
    \includegraphics[trim = 20mm 60mm 35mm 20mm, clip, scale=0.5]{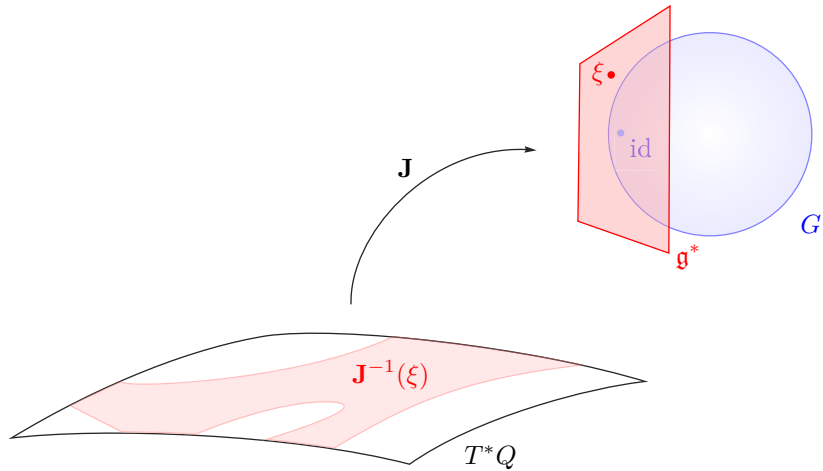}
  \end{center}
  \caption{Symmetries allow to restrict phase space $T^*Q$ to the level sets $\mu^{-1}(\xi)$ of the conserved quantities which are elements $\xi$ in the dual Lie algebra $\g^*$. The connection between phase space and the dual Lie algebra is established by the momentum map $\mu^*$ which provides the modern Hamiltonian formulation of Noether's theorem.}
  \label{fig:math:gm:primer:reduction}
\end{figure}

So far we did not consider symmetries when we described the time evolution of mechanical systems.
Nonetheless, they are a vital aspect of geometric mechanics since they allow to restrict a system's dynamics to the level sets of the conserved quantities, see Fig.~\ref{fig:math:gm:primer:reduction}.
This is possible by Noether's theorem which assures us that continuous symmetries lead to conserved quantities invariant under the dynamics.
We already encountered one such quantity: the Hamiltonian or energy of a system. Conservation of energy is associated with invariance under time translation, and it shows that we tacitly assumed the time invariance of the Hamiltonian in the foregoing.
Other conserved quantities that are often encountered are linear and angular momentum which are associated with translational and rotational invariance, respectively, symmetries we saw before for the classical particle and the pendulum.
Restricting the dynamics of a system to the level sets of the conserved quantities is known as reduction and for many systems critical to obtain an effective description of its time evolution.

\section{Old wine in new skins?}
We have seen that manifolds arise naturally in the description of classical mechanical systems by providing the space of all possible configurations, and that many constraints are enforced intrinsically by the topology and shape of the geometry.
Time evolution in geometric mechanics is represented by curves on the configuration manifold, and we outlined how it can be determined using Lagrangian and Hamiltonian mechanics. 

An important trait of geometric mechanics is its intuitive nature.
The structure and time evolution of a mechanical system can be illustrated by visualizing configuration and phase space, which as we have done on the preceding pages is possible even if the spaces are complex and high dimensional, and practically working with geometric mechanics often means to exploit this inherent geometric intuition.\footnote{The intuitionist approach in geometric mechanics is to be understood in the na{\"i}ve sense of the word, and not in the sense of Brouwer's programme.} 
It goes without saying that formal mathematics has its place in geometric mechanics by complementing intuition and making it rigorous, and that the mastery of modern differential geometry and tensor analysis is essential when geometric mechanics is employed to study mechanical systems. 
Next to the intuitive appeal, a second characteristic of geometric mechanics is its emphasis on mathematical and physical structure. 
While Newtonian mechanics is highly descriptive, making it easy to learn and to carry out computations, it does not reveal structure.
In geometric mechanics, in contrast, computations are structural arguments which provide insight into the fabric they represent, and this structural insight explains much of the vigour of geometric mechanics.\footnote{Lanczos described this as: ``Since the days of antiquity it has been the privilege of the mathematician to engrave his conclusions, expressed in a rarefied and esoteric language, upon the rocks of eternity. While this method is excellent for the codification of mathematical results, it is not so acceptable to the many addicts of mathematics, for whom the science of mathematics is not a logical game, but the language in which the physical universe speaks to us, and whose mastery is inevitable for the comprehension of natural phenomena.'',~\parencite[p. vii]{Lanczos1961}.}

In our discussion we only considered classical mechanical systems.
However, the theory applies to a diverse array of fields and disciplines ranging from quantum mechanics at the smallest scales to relativistic astrophysics at the largest, and applications can be found in areas such as image processing, space mission design, marine animal propulsion, mathematical finance, rising eggs, oceanography, plasma physics, falling cat phenomena, and many more.
In its contemporary formulation using the rich toolbox of modern geometry, geometric mechanics provides thereby a surprisingly unified perspective on all these systems.\footnote{Marsden and Ratiu expressed the unified perspective provided by geometric mechanics as follows: ``Even more striking are true statements like this: 'Don't tell me that quantum mechanics is right and classical mechanics is wrong---after all, quantum mechanics is a special case of classical mechanics.'',~\parencite[p. 116]{Ratiu1999}.}
Although we considered them only cursory, symmetries are an integral part of geometric mechanics and much of the theory is devoted to reduction theory: obtaining simpler descriptions of systems by exploiting symmetries and conserved quantities.

In the next section we will introduce the formal mathematics that is needed for geometric mechanics, and afterwards the intuitive perspective of the theory presented in this section will be made rigorous. 
Nonetheless, even when a more formal language is employed than in our primer, it should be kept in mind that geometric mechanics is an intuitive endeavour and that pictures are the key to understanding.

\subsection*{Acknowledgement}
I would like to thank Tyler de Witt and Eugene Fiume for discussions.

\newpage
\printbibliography[maxnames=20]

\end{document}